\documentclass[twocolumn,showpacs,a4paper,superscriptaddress,nofootinbib,tightenlines,floats]{revtex4} 
\usepackage{bm}
\usepackage{latexsym}
\usepackage{dcolumn}
\usepackage{amsfonts,amssymb}
\usepackage{graphicx}
\usepackage{psfrag}
\usepackage{amsmath}
\begin{document}

\def\apjl{Astrophys. J. Lett.}
\def\mnras{Mon. Not. Roy. Astron. Soc.}
\def\mnrasl{Mon. Not. Roy. Astron. Soc. Lett.}
\def\physrep{Phys. Rept.}
\def\apjs{Astrophys. J. Suppl.}
\def\aap{Astron. Astrophys.}
\def\araa{Annu. Rev. Astron. and Astrophys.}
\def\aj{Astron. J.}

\title{A comparison of perturbations in fluid and scalar field models of dark
  energy }  

\author{H.~K.~Jassal}

\email[$^a$]{Email: hkj@hri.res.in}

\affiliation{Harish-Chandra Research Institute, Chhatnag Road,\\
Jhunsi, Allahabad 211 019, India.\\
}

\begin{abstract}
We compare perturbations in a fluid model of dark energy with those in a scalar
field.  
As compared to the $\Lambda$CDM model,  large scale matter power spectrum is
suppressed in fluid model as well as in a generic quintessence dark energy
model. 
To check the efficacy of fluid description of dark energy in emulating a
scalar field, we consider a potential which gives the same background
evolution as a fluid with a constant equation of state.
We show that for sub-Hubble scales, a fluid model effectively emulates a
scalar field model. 
At larger scales, where dark energy perturbations may play a significant role,
the fluid analogy breaks down and the evolution of matter density contrast
depends on individual scalar field models.  
\end{abstract}


\maketitle

Cosmological observations have confirmed that the expansion of the
universe is accelerating at present \cite{obs_proof}. 
These observations include Supernova type Ia observations
\cite{nova_data},
observations of Cosmic Microwave Background \cite{boomerang,wmap_params} and
large scale structure \cite{sdss}. 
The accelerated expansion of the universe can  be explained by
introducing a cosmological constant $\Lambda$ in the Einstein's equation
\cite{ccprob_wein,review3,varunrev}.
However, the cosmological constant model is plagued by the fine tuning
problem \cite{ccprob_wein}.
This has motivated studies of dark energy models to explain the current
accelerated expansion of the universe \cite{DEreview}.
A typical feature of these models is that the dark energy density varies with
time.

Varying dark energy is typically realized as an ideal fluid or as a scalar
field. 
Purely from distance measurements, it is not possible to distinguish between
different models with the same background evolution.
Evolution of perturbations in these models is expected to break this
degeneracy.
In principle, Integrated Sachs Wolfe (ISW) effect can distinguish a
cosmological constant from other models of dark energy, especially ones with a
dynamical dark energy \cite{ddw}. 
Dark energy perturbations have been extensively studied in the linear
approximation \cite{weller_lewis,bean_dore,depert,gordonhu}.
It was shown in \cite{weller_lewis} that dark energy perturbations affect the
low $l$ quadrupole in the CMB angular power spectrum through the ISW effect.
For models with $w>-1$ this effect leads to an enhancement in power while for
phantom like models it leads to a suppression.
Dark matter perturbations and dark energy perturbations are
anti-correlated for large effective sound speeds. 
This anti-correlation is a frame dependent effect and vanishes if one
considers dark energy rest frame instead of dark matter rest frame
\cite{bean_dore}.
There are several other studies of perturbations in dark energy
\cite{chpgas_pert,sph_coll}, including some that deal with evolution of
spherical perturbations: e.g. see Mota et al\cite{mota}.

In this Brief Report we revisit dark energy perturbations in the context of a
perfect fluid model and compare the evolution with scalar field models. 
To describe dark energy perturbations, we choose the  Newtonian gauge.
In the absence of anisotropic stress, the perturbed metric can be
written in the form   
\begin{equation}
ds^2 = (1+ 2\Phi) dt^2 -  a^{2}(t) \left[(1- 2\Phi)
  \delta_{\alpha \beta} dx^{\alpha} dx^{\beta}\right]
\end{equation}
where $a(t)$ is the scale factor and $\Phi$ is the gauge invariant potential
defined in \cite{bardeen}.  We have assumed the universe to be spatially flat.
The Newtonian potential $\Phi$ characterizes the metric perturbations.

The linearized perturbed Einstein equations for the above metric are given by 
\begin{eqnarray} \label{eq:einstein}
\frac{k^2}{a^2} \Phi + 3 \frac{\dot{a}}{a} \left(\dot{\Phi} +
  \frac{\dot{a}}{a} \Phi \right) &=&  \\ \nonumber
&-& 4\pi G \left[\rho_{NR} \delta_{NR} + \rho_{DE} \delta_{DE} \right] \\
\nonumber
\dot{\Phi} + \frac{\dot{a}}{a} \Phi &=& -4 \pi G  \left[\rho_{NR} v_{NR} +
  \rho_{DE} v_{DE} \right] \\
\nonumber 
4 \frac{\dot{a}}{a} \dot{\Phi} + 2  \frac{\ddot{a}}{a} \Phi +
\frac{\dot{a}^2}{a^2} \Phi + \ddot{\Phi} &=& 4 \pi G c_s^2 \rho_{DE} \delta_{DE} 
\end{eqnarray}
where dot denotes derivative with respect to the coordinate time $t$, with
$\rho_{NR}$ and $\rho_{DE}$ representing energy density for nonrelativistic
and dark energy components and $P$ denotes pressure. 
The comoving velocity of a fluid or a field is denoted by $v$. 
Here $\delta_{NR} \equiv \delta \rho_{NR} / \rho_{NR}$ is the density contrast
for nonrelativistic matter and $\delta_{DE}$ is the density contrast for dark 
energy. The symbol $c_s^2 \equiv \delta P/\delta \rho$ denotes the speed of
propagation of perturbations.

We first describe a fluid model of dark energy.
For a fluid with a constant equation of state parameter $w \equiv P/\rho$, the
continuity equation and Euler equation reduce to 
\begin{eqnarray} \label{eq:cont_euler}
\dot{\delta} &=& (1 + w) \left[-\nabla_{\alpha}v^{\alpha} + 3 \dot{\Phi} - 3
  \frac{\dot{a}}{a} (c_s^2 -w) \delta
  \right] \\ \nonumber
\dot{v} &=& \Phi + \frac{c_s^2}{1+w} \delta + \frac{\dot{a}}{a}\left[3w-1\right]v.
\end{eqnarray}

It is useful to Fourier transform the  above equations as in the linear regime
the modes evolve independent of each other.
Since we will discuss Fourier modes in the rest of the paper we do not use the
subscript $k$.

\begin{figure}[t]
\begin{center}
\includegraphics[width=2.5in]{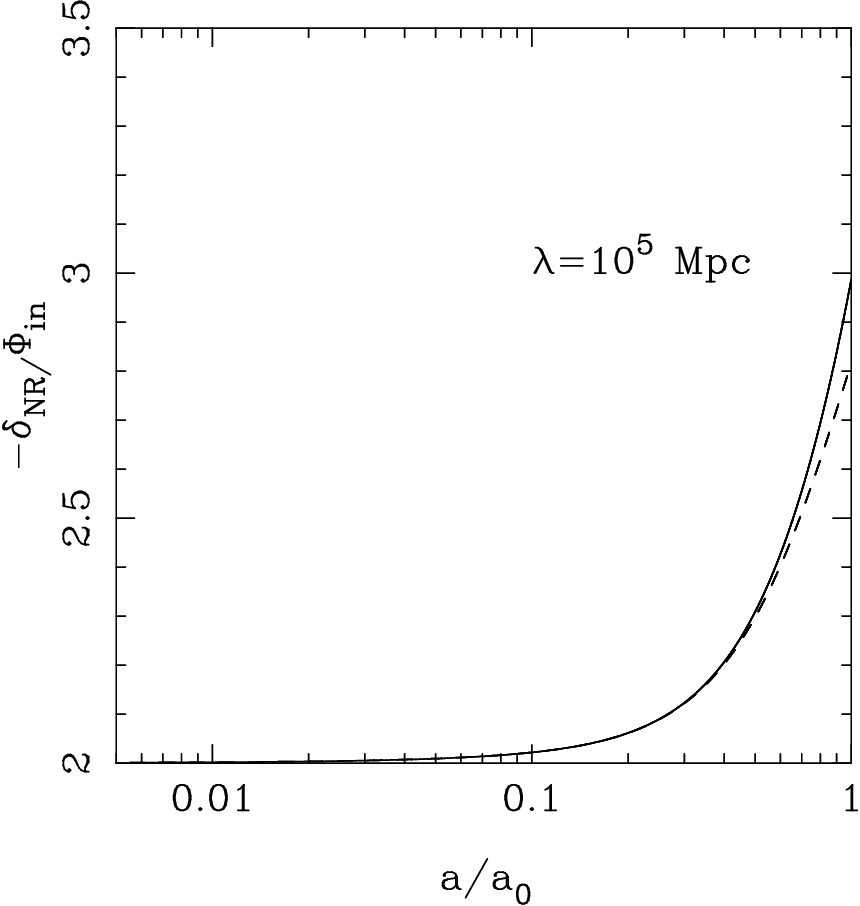}
\end{center}
\caption{The figure shows the evolution of density contrast for
  nonrelativistic matter (scaled by initial gravitational potential
  $\Phi_{in}$)  at   $\lambda=10^5$ Mpc   for the fluid model with
  $w=-0.8$. The   solid line is   evolution if     we assume dark energy to be
  a smooth   component of the   universe. The dashed   line is  evolution of
  $\delta_{NR}$ if dark energy   clusters.}   
\label{fig::fluiddm}
\end{figure}

For our discussion, we choose the following two equations
\begin{eqnarray}
\label{eq:maineqs}
\Phi''  + 4 \frac{\dot{a}}{a} \Phi' &=& 4 \pi G \rho_{DE}
(c_s^2 \delta_{DE}  + 2 w \Phi) \\ \nonumber
\delta''_{DE} &+& (2 + 2 c_S^2- 6w)  \frac{\dot{a}}{a} \delta'_{DE} + c_s^2 \frac{\bar{k}^2}{a^2} \delta_{DE}  \\ \nonumber
&-&(c_s^2 - w) \left[\frac{3}{2}\frac{a'^2}{a^2} + \frac{3w}{2} \frac{\Omega_{DE}}{a^{3(1+w)}} + 9 w \frac{a'^2}{a^2}  \right] \delta_{DE}
 \\ \nonumber
&=& (1+w) \left[-\frac{\bar{k}^2}{a^2} \Phi + 3(2 - 3w)\frac{a'}{a} \Phi'
  +3 \Phi'' \right]  
\end{eqnarray}
We obtain the second equation by differentiating the first equation in (3) and
eliminating $v$.
The present day Hubble parameter is denoted by $H_0$ and $\Omega_{DE}$ is the
density parameter for the dark energy component.
The prime denotes derivative with respect to variable $x=tH_0$ and
$\bar{k}=kc/H_0$. 

\begin{figure}[t]
\begin{center}
\includegraphics[width=2.5in]{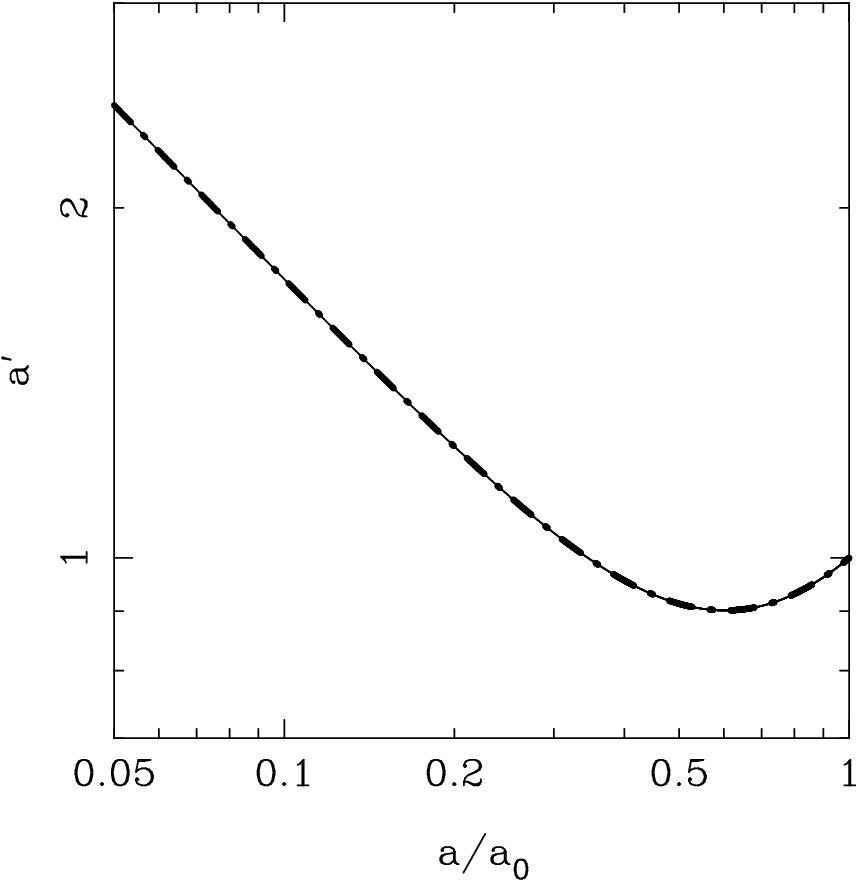}
\end{center}
\caption{The figure shows the evolution of $a'$ as a function of $a$. The
  solid line is for a fluid model with $w=-0.86$ and the dot dashed line is
  for the model with 'reconstructed' potential.} 
\label{fig::bg}
\end{figure}

\begin{figure*}
\begin{center}
\begin{tabular}{cc}
\includegraphics[width=2.5in]{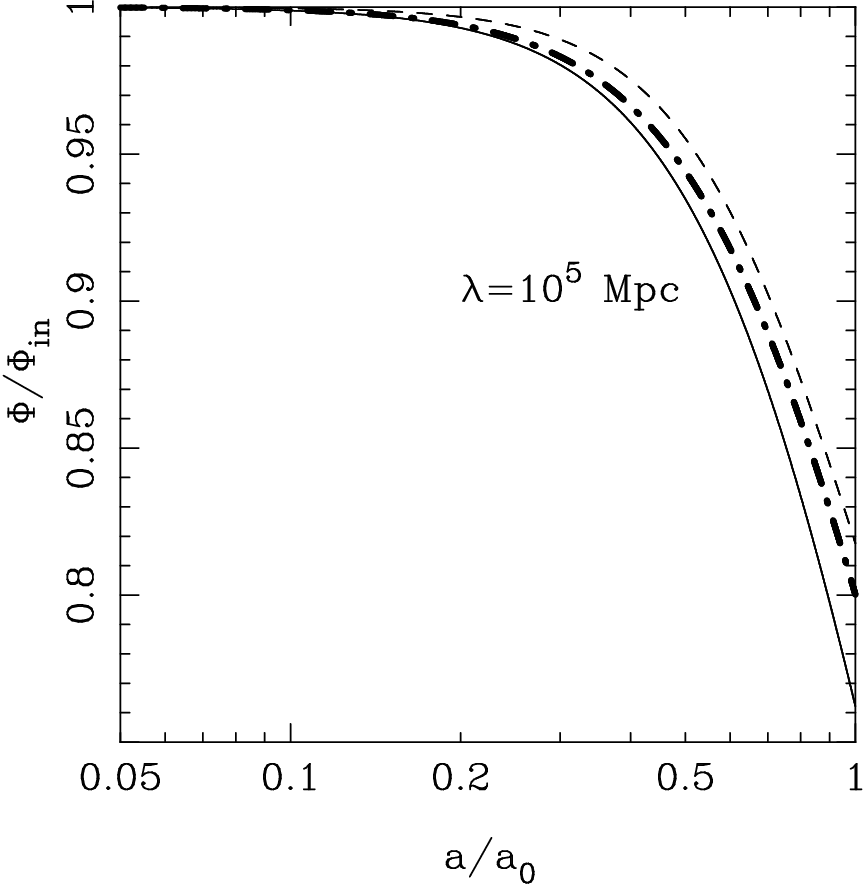}
\includegraphics[width=2.5in]{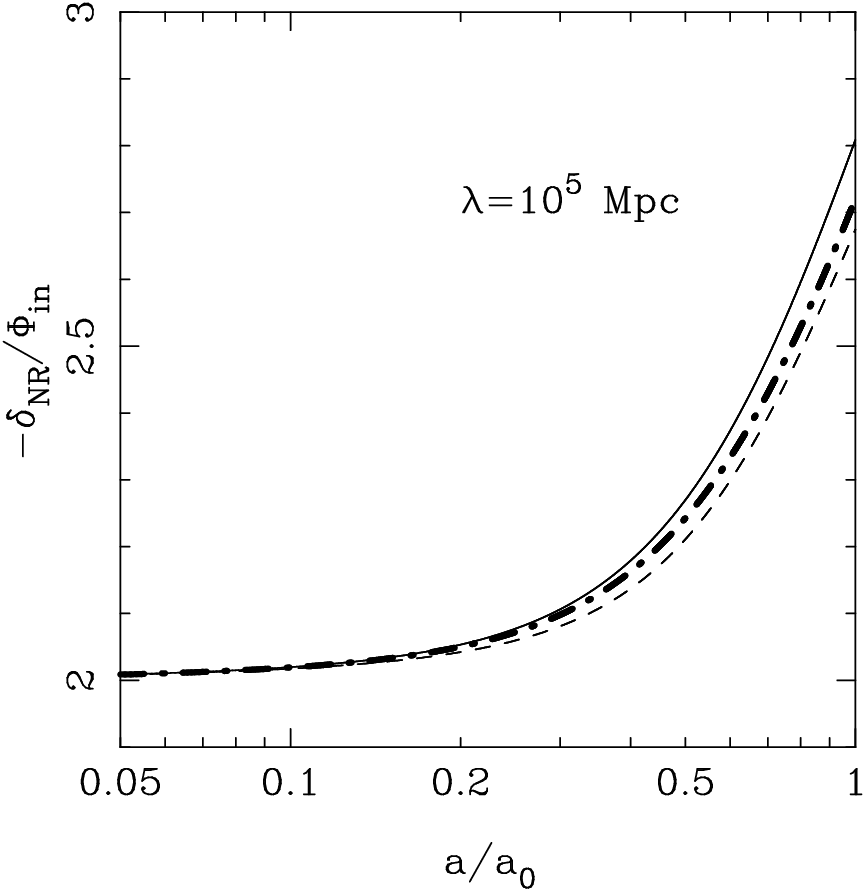}
\end{tabular}
\end{center}
\caption{The figure on the left shows the evolution of the gravitational
  potential $\Phi$ as a function of   $a$ and the one on right shows matter
  density contrast for fluid model   (solid line), for exponential potential
  (dashed   line) and for the   'reconstructed' potential (thick dot dashed
  line) at $\lambda=10^5$ Mpc.}    
\label{fig::compare}
\end{figure*}

In the first equation in system (\ref{eq:maineqs}), the $\delta_{DE}$ term on
the right hand side is the main departure from the concordance $\Lambda$CDM
cosmology.  
If dark energy is a cosmological constant, then it does not cluster and the
field $\Phi$ decays whenever $\Lambda$ dominates. 
For $w \neq -1$, the potential $\Phi$ remains constant in the matter dominated
era and starts to decay when dark energy contribution becomes significant. 
If $c_s^2=w$, as soon as the dark energy component dominates, the effect of
$\delta_{DE}$ term is to increase the potential $\Phi$. 
This is a scale dependent effect and is more pronounced at small scales, i.e.,
for large $k$ values.
This increase is due to the fact that as dark energy starts to dominate,
gradient in pressure enhances potential gradient and in turn dark energy
clusters faster than matter \cite{gordonhu}. 
For a positive sound speed, this instability is not there and the gravitational
potential continues to decay in the dark energy dominated era.

In Fig. \ref{fig::fluiddm}  we plot evolution of $\delta_{NR}$ as a function
of the scale factor $a$ at length scale $\lambda=10^5$ Mpc. 
In this figure we fix the parameter $c_S^2=1$.
At small scales, i.e., at length scales smaller than the Hubble radius, the
evolution of the matter density contrast in perturbed dark energy scenario
matches the evolution if dark energy is assumed to be homogeneous: due to a
positive sound speed $\delta_{DE}$ has an oscillatory behavior and these
perturbations do not grow and hence matter perturbations behave as if dark
energy is homogeneous.  
At large scales, as soon as dark energy starts to dominate, the difference
starts to increase.
Varying $c_s^2$ does not make any difference as long as it is positive.
In such a case there is no instability by way of a growing $\Phi$.
The evolution of $\Phi$ is the same for all $\lambda < $ Hubble radius. 
The gravitational potential remains constant in the matter dominated era and
starts to decay as dark energy dominates.
At scales larger than the Hubble radius the decay of potential is
slower.  
For models with $w>-1$, there is a correlation in dark energy density contrast
and the matter density contrast. 
For 'phantom' models, these are anticorrelated, i.e. an overdensity in matter
corresponds to an underdensity in dark energy.
We will however consider only 'quintessence' type fields in this Report.

To differentiate between the role of background evolution and that of dark
energy perturbations on matter perturbations, we choose a scalar field
potential which emulates a constant dark energy equation of state.
We call it the 'reconstructed' potential for further discussion.
To emulate the background evolution corresponding to a constant equation of
state, the scalar field potential is given by \cite{varunrev,affine}
\begin{equation}
V(a) = \frac{3}{2} (1-w)  H_0^2  M_P^2 \frac{\Omega_{DE}}{a^{3(1+w)}}
\end{equation}
with evolution of scalar field $\varphi$ given by
\begin{equation}
\frac{d\varphi}{da}= \frac{\sqrt{3(1+w)\Omega_{DE}M_P^2}}
{a^{(5+3w)/2}\sqrt{\Omega_{NR}a^{-3} + \Omega_{DE}a^{-3(1+w)}}}
\end{equation}

This potential gives the same background solution as that of a fluid with a
constant $w$ (as can be seen in Fig. \ref{fig::bg}).
In the dark energy dominated era, it can be checked that the potential is
an exponential one. 
We substitute the above in the perturbation equations for a scalar field
\begin{equation}
\delta \varphi'' + 3\frac{a'}{a} \delta \varphi' + \frac{\bar{k}^2
  \delta 
  \varphi}{a^2} + 2\Phi \frac{dV(\varphi)}{d \varphi}
-4 \Phi' \varphi' + \frac{d^2 V(\varphi)}{d \varphi^2}\delta \varphi = 0
\end{equation}
and the linearized Einstein equation
\begin{eqnarray}
\label{eq::ptnl}
\Phi'' + 4\frac{a'}{a}\Phi' + \big (2 \frac{a''}{a} +
\frac{a'^2}{a^2} \big)\Phi &=& 4\pi G \left[ \varphi' \delta\varphi' -\Phi
  \varphi'^2 \right. \\ \nonumber
&-&  \left. \frac{d V(\varphi)}{d \varphi} \delta \varphi \right] 
\end{eqnarray}

In the fluid model of dark energy, matter perturbations are suppressed as
compared to the homogeneous dark energy model (this agrees with the result of
\cite{sph_coll} in the linear limit). 
This suppression is in contrast to the (quintessence) scalar field models
where matter perturbations are enhanced by the presence of dark energy
perturbations \cite{ujs}. 
This difference is due to the fact the homogeneous limit is achieved
differently for the two models. 
For a scalar field,  even if one ignores spatial gradients there still remains
a contribution to $\delta P$. 
This contribution cancels with a corresponding contribution from the pressure
term due to the background i.e., from the third term on the left hand side of
Eq. \ref{eq::ptnl}. 
The cancellation of this term, for smooth dark energy, leads to a suppression
in matter perturbations.
This suppression is due to the assumption that dark energy is a smooth
component. 
Therefore one must take care while taking this limit of quintessence models as
the smooth limit is inconsistent.
In other words, if dark energy is allowed to cluster the matter perturbations
are enhanced in comparison.
For fluid models $\delta P$  vanishes and the residual pressure term due to
the background evolution makes the evolution different from that in a scalar
field. 
Therefore, assuming dark energy to be homogeneous leads to a large difference
in matter density contrast.
Perturbations increase the degeneracy between different models.
As compared to the $\Lambda$CDM matter perturbations are  suppressed  
for both fluid and for scalar field models.

We now  compare the evolution of perturbations in the 'reconstructed' model
with that in a fluid model. 
The fluid approximation works very well for sub-Hubble scales. 
The evolution of $\delta_{NR}$ in this scalar field model is identical to the
evolution in fluid model (with $c_S^2=1)$. 
This approximation breaks down at scales above the Hubble radius. 
The effects of the scalar field potential is dominant over that of the sound
speed and $\delta_{NR}$ closely follows that in the exponential potential. 

In Fig. \ref{fig::compare} we show the gravitational potential $\Phi$ and
$\delta_{NR}$ as functions of $a$ for fluid model, exponential potential
$V(\varphi)=V_0 exp[-\sqrt{\alpha} \varphi/M_P]$ \cite{ujs} ($M_P$ being
Planck mass), a fluid model with $w=-0.86$ and the corresponding
'reconstructed' potential at $\lambda=10^5$ Mpc. 
For the exponential potential, if $\alpha =1$, the dark energy equation of
state $w \approx -0.86$. 
For small scales, say for $\lambda=50$ Mpc, the evolution in all three cases
is very similar. 
At the present epoch, the percentage difference in the fluid model and the
reconstructed potential model is less than $10^{-5}$\%. 
The difference between the fluid model and the exponential potential model is
$\approx 0.14$\%.
The reconstructed potential and fluid model start to differ on Hubble scale
and larger.
This difference remains less than $1$\% upto the scale $\lambda=10^4$.
For larger scales, i.e., $\lambda=10^5$ Mpc, as soon as dark energy starts to
dominate, the behavior of reconstructed potential model becomes closer to
that of an exponential potential.  
The percentage difference in the fluid model and the reconstructed potential
model is $\approx 3.6$\% and the difference between the reconstructed
potential and the exponential potential model is $\approx 0.8$\%.
The dominant term in time evolution of $\delta_{NR}$ in both the cases is
$\dot{a}^2 \Phi/a^2$.
The other terms suppress this evolution and these terms are model dependent
with the scalar field potential and its derivatives having a more pronounced
effect in this opposition than in the fluid model.
At large scales therefore, in the dark energy dominated epoch, the properties
of the scalar field potential come into play.

The results of this work are summarized as follows.
Dark energy perturbations in fluid models suppress matter perturbations as
compared to the corresponding smooth dark energy model.
Matter perturbations are suppressed as compared to the $\Lambda$CDM model. 
As long as the sound speed is positive the evolution of matter
perturbations is indistinguishable from a smooth dark energy model. 
This is true for scales smaller then the Hubble radius.
Dark energy perturbations in a fluid model emulate that of a scalar
field model very well below the Hubble scale but start to differ at larger
scales.
Therefore the fluid model is not a good approximation at these scales.
This also implies that the growth of perturbations at large scales depends on
the details of the model even though the background evolution is the
same. 
A separate analysis is therefore required for every  model.

The author is grateful to J. S. Bagla, T. Padmanabhan,
T. R. Seshadri, K. Subramanian and S. Unnikrishnan for useful discussion.  
The author thanks Department of Science and Technology, India for financial
assistance through project number SR/WOS-A/PS-11/2006.

\end{document}